\documentclass{ptephy_v1}

\begin{document}

\Year{2022}
\artid{023A01}
\setDOI{ptab165}{}

\makeatletter \@addtoreset{equation}{section} \makeatother
\def\theequation{\arabic{section}.\arabic{equation}}

\newcommand{\myset}[1]{\mathrm{#1}}
\renewcommand\b{b}
\renewcommand\k{k}
\newcommand\myvar[2]{\stackrel{\scriptscriptstyle #2}{#1}}
\newcommand\bu[1]{\myvar{u}{#1}}
\newcommand\ku[1]{\myvar{v}{#1}}
\newcommand\dm[1]{\gamma^{#1}}

\newcommand\eref[1]{(\ref{#1})}
\newcommand\dirac{\not{\!\partial}}
\newcommand\Lagrangian{\mathcal{L}}
\newcommand\hconj{^{\dagger}}
\newcommand\Ns{N_{\!s}}
\newcommand\Nz{N_{\!f}}
\newcommand\transposed{^{\scriptscriptstyle{T}}}
\renewcommand\vec[1]{\boldsymbol{#1}}

\title{Dark solitons of the Gross-Neveu model.}

\author{V.E. Vekslerchik} 

\affil{
  Usikov Institute for Radiophysics and Electronics, \\
  12, Proskura st., Kharkov, 61085, Ukraine 
  \email: \texttt{vekslerchik@yahoo.com}
}

\subjectindex{A10,A11,B34}

\begin{abstract}%
We present N-soliton solutions for the classical (1+1)-dimensional Gross–Neveu 
model which satisfy non-zero boundary conditions. These solutions are obtained 
by direct method using some properties of the soliton matrices that appear in 
the framework of the Cauchy matrix approach.
\end{abstract}

\maketitle

\section{Introduction.}

In this paper we consider the (1+1)-dimensional Gross–Neveu model with 
$\Nz$ flavors of Dirac fermions \cite{GN74},
\begin{equation} 
\label{def:LagrGN}
  \mathcal{L} 
  = 
  i \sum_{n=1}^{\Nz} \bar{\vec\psi}_{n} \!\dirac \vec\psi_{n}  
  + g \biggl( \sum_{n=1}^{\Nz} \bar{\vec\psi}_{n} \vec\psi_{n} \biggr)^{2}
\end{equation}
(we will explain all the designations in what follows).
This two-dimensional massless fermion asymptotically free field model 
has been introduced in 1974 in connection with the search for symmetry breaking, 
and since then has attracted a lot of interest in semiclassical field theory. 
The solutions of classical equations corresponding to the Lagrangian $\mathcal{L}$ 
may be considered as candidates, or classical approximations, for the particles 
of the corresponding quantum theory. 
That is why the analytic solution of the classical model remains an actual 
problem (see, for example, 
\cite{F95,F04,T04,KT10,FT11,TTYN12,TN13,TN14,DT13a,DT13b,T14,DT14,T20}). 

At the classical level, the Gross-Neveu model is closely related to the theory 
of integrable systems. In papers \cite{NP78,ZM80} the authors found a 
class of the Gross-Neveu-like models which are completely integrable, and one 
can find there the inverse scattering transform which gives the possibility 
of deriving various solutions, in particular the soliton ones. 
Looking at more recent 
works \cite{KT10,FT11,TTYN12,TN13,TN14,DT13a,DT13b,T14,DT14,T20} one can notice that, 
although the authors do not use the results of, say, \cite{ZM80} directly, they use 
various approaches developed in the theory of integrable systems: 
the Zakharov-Shabat scattering problem, Gelfand-Levitan-Marchenko equations 
(in \cite{TTYN12,TN13,TN14}), or the theory of reflectionless potentials 
\cite{KM56,DHN74a,DHN74b}, Hirota ansatz and inverse scattering transform for 
the $\sinh$-Gordon equation (in \cite{KT10,FT11,DT13a,DT13b,T14,DT14,T20}). 

The main object of this work is the $\Ns$-soliton solutions for the classical 
Gross-Neveu model \eref{def:LagrGN} or, in matrix form, 
\begin{equation}
\label{def:LagrGNa} 
  \Lagrangian
  = 
  i \mbox{Tr}\; \mathsf{\bar{\Psi}} \!\dirac\mathsf{\Psi} 
  + g \left(\mbox{Tr}\; \mathsf{\bar{\Psi}} \mathsf{\Psi} \right)^{2} 
\end{equation}
where $\mathsf{\Psi} = \left( \vec\psi_{1}, \, ... \, \vec\psi_{\Nz} \right)$, 
which were derived in the paper \cite{FT11} by Fitzner and Thies. 
Thus, the essential part of this paper may be viewed as an alternative 
derivation and representation of the results of \cite{FT11}. 
The method used in what follows is a variant of the 
Cauchy matrix approach \cite{NQC83,QNCL84,NAH09,ZZ13,HJN16,FN17}. 
We start with an ansatz based on some class of matrices, the so-called 
`almost-intertwining' matrices \cite{KG01} that satisfy the `rank one 
condition' \cite{GK02,GK06a,GK06b}, which is a particular case of the 
Sylvester equation \cite{DM10a, DM10b, XZZ14, SZN17}. 
An analysis of the properties of these matrices leads us to solutions of the 
`two-field' model 
\begin{equation}
\label{def:LagrGNb} 
  \Lagrangian
  = 
  i \mbox{Tr}\; \mathsf{\Phi} \!\dirac\mathsf{\Psi} 
  + g \left(\mbox{Tr}\; \mathsf{\Phi} \mathsf{\Psi} \right)^{2}
\end{equation}
(see section \ref{sec:auxiliary}).
The important point is that we do not need to solve the 
`consistency' equations separately. We just show in sections 
\ref{sec:constraints} and \ref{sec:involution} that the proposed ansatz 
possesses some \emph{reduction}, that automatically resolves the consistency 
condition and leads from \eref{def:LagrGNb} to \eref{def:LagrGNa}.  
This gives the possibility of obtaining the Gross-Neveu solitons, 
which we discuss in section~\ref{sec:solitons}.

\section{Auxiliary system. \label{sec:auxiliary}}

\subsection{Almost-intertwining matrices.}

The solutions we present here are built of the matrices defined by the equation
\begin{equation}
    \mathsf{L} \mathsf{A}
  - \mathsf{A} \mathsf{R}
  = 
    | \k \rangle \langle \b |.
\label{eq:sy}
\end{equation}
Here, 
$\mathsf{L}$ and $\mathsf{R}$ are constant diagonal $\Ns\times\Ns$ matrices, 
$\langle \b |$ is $\Ns$-component row and   
$| \k \rangle$ is $\Ns$-component column, 
\begin{equation}
\label{def:bk}
  \langle \b | = \left( b_{1}, \, ... \, , b_{\Ns} \right), \qquad
  | \k \rangle = \left( k_{1}, \, ... \, , k_{\Ns} \right)\transposed.
\end{equation}
The dependence of 
$\langle \b |$ and $| \k \rangle$ on the coordinates describing the model, 
$\langle \b | = \langle \b(\xi,\eta) |$ and 
$| \k \rangle = | \k(\xi,\eta) \rangle$, 
is defined by 
\begin{equation}
\label{def:doib}
  \partial_{\xi}\langle \b | = \langle \b | \mathsf{R}^{-1},
  \qquad
  \partial_{\eta}\langle \b | = \langle \b | \mathsf{R}
\end{equation}
and 
\begin{equation}
\label{def:doik}
  \partial_{\xi}| \k \rangle = - \mathsf{L}^{-1} | \k \rangle,
  \qquad
  \partial_{\eta}| \k \rangle = - \mathsf{L} | \k \rangle
\end{equation}
which leads to 
\begin{equation}
  \partial_{\xi}\mathsf{A} = | \k_{o} \rangle \langle \b_{o} |, 
  \qquad
  \partial_{\eta}\mathsf{A} = - | \k \rangle \langle \b |
\end{equation}
where 
\begin{equation}
\label{def:boko}
  \langle \b_{o} | = \langle \b | \mathsf{R}^{-1}, 
  \qquad
  | \k_{o} \rangle = \mathsf{L}^{-1} | \k \rangle.
\end{equation}

In what follows, we use another set of rows and columns, this time of the 
length $\Nz$, defined by 
\begin{equation}
\label{def:bu}
  \begin{array}{l} 
  \langle \bu1_{\myset{Z}} |
  = 
    \langle 1_{\myset{Z}} |
  - \langle \b | \mathsf{G} \mathsf{K}_{\myset{Z}},
  \\[2mm]
  \langle \bu2_{\myset{Z}} |
  = 
    \langle 1_{\myset{Z}} | \mathsf{D}_{\myset{Z}}^{-1}
  - \langle \b_{o} | \mathsf{G} \mathsf{K}_{\myset{Z}}
  \end{array} 
\end{equation}
and 
\begin{equation}
\label{def:ku}
  \begin{array}{l} 
  | \ku1_{\myset{Z}} \rangle
  = 
  | 1_{\myset{Z}} \rangle
  + \mathsf{B}_{\myset{Z}} \mathsf{G} | \k \rangle,
  \\[2mm]
  | \ku2_{\myset{Z}} \rangle
  = 
  \mathsf{D}_{\myset{Z}}^{-1} | 1_{\myset{Z}} \rangle
  + \mathsf{B}_{\myset{Z}} \mathsf{G} | \k_{o} \rangle.
  \end{array} 
\end{equation}
Here, 
\begin{equation}
  \mathsf{G} = \left( \mathsf{1} + \mathsf{A} \right)^{-1},
\end{equation}
$\myset{Z}$ is a $\Nz$-set of constants 
$\myset{Z} = \{z_{1}, \, ... \, , z_{\Nz}\}$, 
$\langle 1_{\myset{Z}} |$ and $| 1_{\myset{Z}} \rangle$ 
are $\Nz$-row and $\Nz$-column with all components equal to $1$, 
and $\mathsf{D}_{\myset{Z}}$ is the diagonal $\Nz\times\Nz$-matrix, 
\begin{equation}
  \mathsf{D}_{\myset{Z}} = 
  \mbox{diag}\left( z_{1}, \, ... \, , z_{\Nz} \right), 
\end{equation}
while 
$\mathsf{B}_{\myset{Z}}$ and $\mathsf{K}_{\myset{Z}}$ 
are rectangular matrices given by 
\begin{equation}
  \mathsf{B}_{\myset{Z}} 
  = 
  \Biggl(
    \frac{ b_{n} }{ R_{n} - z_{m} }
  \Biggr)_{\!\!\begin{array}{l} 
           \scriptscriptstyle m = 1, ..., \Nz \\[-2mm] 
           \scriptscriptstyle n = 1, ..., \Ns 
           \end{array} },
\qquad
  \mathsf{K}_{\myset{Z}} 
  = 
  \Biggl(
    \frac{ k_{m} }{ L_{m} - z_{n} }
  \Biggr)_{\!\!\begin{array}{l} 
           \scriptscriptstyle m = 1, ..., \Ns \\[-2mm] 
           \scriptscriptstyle n = 1, ..., \Nz 
           \end{array} }
\end{equation}
where $b_{n}$ and $k_{n}$ are components of 
$\langle \b |$ and $| \k \rangle$ (see \eref{def:bk}).

Applying the rules 
\eref{def:doib} and \eref{def:doik} to the definitions 
\eref{def:bu} and \eref{def:ku} one can obtain, by simple algebra, the 
following equations governing the `evolution' of 
$\langle \bu{i}_{\myset{Z}} |$ and $| \ku{i}_{\myset{Z}} \rangle$:
\begin{equation} 
\label{eq:doibu}
  \begin{array}{l} 
  \partial_{\xi}\langle \bu1_{\myset{Z}}| 
  = 
  - \langle \bu1_{\myset{Z}} | \mathsf{D}_{\myset{Z}}^{-1}
  + u_{o} \langle \bu2_{\myset{Z}} |,
  \\
  \partial_{\eta}\langle \bu2_{\myset{Z}}| 
  = 
  - \langle \bu2_{\myset{Z}} | \mathsf{D}_{\myset{Z}}
  + v_{o} \langle \bu1_{\myset{Z}} |
  \end{array} 
\end{equation}
and 
\begin{equation}
\label{eq:doiku}
  \begin{array}{l} 
  \partial_{\xi}| \ku1_{\myset{Z}}\rangle 
  = 
  \mathsf{D}_{\myset{Z}}^{-1} | \ku1_{\myset{Z}} \rangle
  - v_{o} | \ku2_{\myset{Z}} \rangle,
  \\
  \partial_{\eta}| \ku2_{\myset{Z}}\rangle 
  = 
  \mathsf{D}_{\myset{Z}} | \ku2_{\myset{Z}} \rangle
  - u_{o} | \ku1_{\myset{Z}} \rangle, 
  \end{array} 
\end{equation}
where the functions $u_{o}$ and $v_{o}$ are defined by 
\begin{equation}
\label{def:uv}
  u_{o}
  = 
  1 - \langle \b | \mathsf{G} | \k_{o} \rangle, 
\qquad
  v_{o}
  = 
  1 + \langle \b_{o} | \mathsf{G} | \k \rangle. 
\end{equation}

\subsection{Constraints. \label{sec:constraints}}

The \emph{linear} (with respect to 
$\langle \bu{i}_{\myset{Z}} |$ and $|\ku{i}_{\myset{Z}} \rangle$) 
equations presented in the previous subsection are an important part of the 
approach of this paper. However, now we face a more difficult problem: we have 
to close system \eref{eq:doibu}--\eref{def:uv} (note that there is no obvious 
relationships between $\langle \bu{i}_{\myset{Z}} |$, 
$|\ku{i}_{\myset{Z}} \rangle$, $u_{o}$ and $v_{o}$).
Contrary to the derivation of \eref{eq:doibu} and \eref{eq:doiku}, which is a 
straightforward procedure similar to one used repeatedly by various authors, 
the `closure' problem is less trivial. 
In the framework of the theory of integrable systems, it is related to the 
so-called Bargmann constraints or the nonlinearization procedure. 
We do not discuss here the `theoretical' aspects of this problem. Instead, 
we demonstrate that it is possible to relate the rows 
$\langle \bu{i}_{\myset{Z}} |$ and the columns $| \ku{i}_{\myset{Z}} \rangle$ 
to the functions $u_{o}$ and $v_{o}$ 
by careful analysis of the structure of 
the $\mathsf{A}$-matrices or, in other words, of consequences of equation 
\eref{eq:sy}. 

It can be shown (see Appendix \ref{app:products}) that equation \eref{eq:sy}, 
together with the definitions \eref{def:bu}, \eref{def:ku} and \eref{def:uv}, 
imply 
\begin{equation}
\label{eq:sp}
  \begin{array}{l}
  \langle \bu1_{\myset{Z}} | \ku2_{\myset{Z}} \rangle
  = 
  h u_{o} - \langle \b | \mathsf{F}_{\myset{Z}} | \k_{o} \rangle, 
  \\
  \langle \bu2_{\myset{Z}} | \ku1_{\myset{Z}} \rangle
  = 
  h v_{o} - \langle \b_{o} | \mathsf{F}_{\myset{Z}} | \k \rangle
  \end{array}
\end{equation}
where 
\begin{equation}
\label{def:h}
  h = \sum_{n=1}^{\Nz} z_{n}^{-1}, 
\end{equation}
and 
\begin{equation} 
\label{def:F}
  \mathsf{F}_{\myset{Z}} = \mathsf{G} \mathsf{\hat{F}}_{\myset{Z}} \mathsf{G}   
\end{equation} 
with 
\begin{equation}
  \mathsf{\hat{F}}_{Z} 
  = 
  \mathop{\mbox{diag}}
  \left( 
  \sum_{n=1}^{\Nz} 
    \left[ 
    \frac{ 1 }{ L_{s} - z_{n} } 
    - 
    \frac{ 1 }{ R_{s} - z_{n} }
    \right]
  \right)_{s = 1, \, ... \, , \Ns} 
\end{equation}
(here, $L_{s}$ and $R_{s}$ are the elements of the diagonal matrices 
$\mathsf{L}$ and $\mathsf{R}$).

As one can see from equations \eref{eq:sp}, the variables 
$\langle \bu{i}_{\myset{Z}} |$, $|\ku{i}_{\myset{Z}} \rangle$, 
$u_{o}$ and $v_{o}$ which are involved in equations 
\eref{eq:doibu} and \eref{eq:doiku} are not enough, in the general case, 
to obtain a closed system. 
However, there exists a \emph{reduction} of \eref{eq:sy} that 
eliminates these difficulties. 
 
The key point in our calculations is the fact (demonstrated in  Appendix 
\ref{app:reduction}) that the restriction 
\begin{equation} 
\label{eq:LR}
  \mathsf{L} + \mathsf{R} = 0 
\end{equation} 
leads to the following result:
\begin{equation}
\label{inv:uv}
    u_{o}
  = 
    v_{o}
\end{equation}
and 
\begin{equation}
\label{inv:FZ}
  0
  = 
    \langle \b | \mathsf{F}_{\myset{Z}} | \k_{o} \rangle 
  + \langle \b_{o} | \mathsf{F}_{\myset{Z}} | \k \rangle, 
\end{equation}
which together with \eref{eq:sp} implies 
\begin{equation}
\label{eq:constr}
    \langle \bu1_{\myset{Z}} | \ku2_{\myset{Z}} \rangle 
  + \langle \bu2_{\myset{Z}} | \ku1_{\myset{Z}} \rangle 
  = 
   2 h u_{o}.
\end{equation}
Thus, the restriction \eref{eq:LR} converts equations 
\eref{eq:doibu} and \eref{eq:doiku} into a closed system 
for $\langle \bu{i}_{\myset{Z}} |$, $|\ku{i}_{\myset{Z}} \rangle$.

\subsection{Matrices $\mathsf{\Phi}$ and $\mathsf{\Psi}$.}

Now, we rewrite equations \eref{eq:doibu}, \eref{eq:doiku} and \eref{eq:constr} 
in a matrix form. Consider the $\Nz\times 2$ matrix 
\begin{equation}
  \mathsf{\Phi}
  = 
  \mathsf{E}^{-1} 
  \left( | \ku1_{\myset{Z}} \rangle \; | \ku2_{\myset{Z}} \rangle \right), 
\end{equation}
and the $2\times\Nz$ matrix 
\begin{equation}
  \mathsf{\Psi}
  = 
  \left( \begin{array}{l} 
    \langle \bu1_{\myset{Z}} | \\ \langle \bu2_{\myset{Z}} | 
  \end{array}\right) \mathsf{E},
\end{equation}
where the $\Nz\times\Nz$ diagonal matrix $\mathsf{E}$, which satisfies 
\begin{equation}
  \partial_{\xi}\mathsf{E} = \mathsf{D}_{\myset{Z}}^{-1} \mathsf{E}, 
  \qquad
  \partial_{\eta}\mathsf{E} = \mathsf{D}_{\myset{Z}} \mathsf{E},
\end{equation}
is introduced to take into account the terms proportional to 
$\mathsf{D}_{\myset{Z}}^{\pm 1}$ in \eref{eq:doibu} and \eref{eq:doiku}.

In terms of $\mathsf{\Phi}$ and $\mathsf{\Psi}$, 
equations \eref{eq:doibu} and \eref{eq:doiku} can be written as 
\begin{equation}
  \phantom{-}
    \mathsf{X}_{\xi} \partial_{\xi}\mathsf{\Psi}
  + \mathsf{X}_{\eta} \partial_{\eta}\mathsf{\Psi} 
  = 
  u_{o} \mathsf{\sigma}_{1} \mathsf{\Psi}, 
\end{equation}
\begin{equation}
  - (\partial_{\xi}\mathsf{\Phi}) \mathsf{X}_{\xi}
  - (\partial_{\eta}\mathsf{\Phi}) \mathsf{X}_{\eta}
  = 
  u_{o} \mathsf{\Phi} \mathsf{\sigma}_{1}, 
\end{equation}
where 
\begin{equation} 
  \mathsf{X}_{\xi} = 
  \left( \begin{array}{cc} 1 & 0 \\ 0 & 0 \end{array} \right), 
  \qquad
  \mathsf{X}_{\eta} = 
  \left( \begin{array}{cc} 0 & 0 \\ 0 & 1 \end{array} \right)   
\end{equation} 
and $\mathsf{\sigma}_{1}$ is the Pauli matrix,
while equation \eref{eq:constr} takes the form 
\begin{equation}
   2 h u_{o}
  = 
  \mbox{Tr}\; \mathsf{\Phi} \mathsf{\sigma}_{1} \mathsf{\Psi}. 
\end{equation}

To summarize, matrices $\mathsf{\Phi}$ and $\mathsf{\Psi}$ satisfy equations 
corresponding to the Lagrangian
\begin{equation}
\label{def:lagrA}
  \Lagrangian
  = 
  4 h \, \mbox{Tr}\; \mathsf{\Phi}\dirac\mathsf{\Psi} 
- \left( \mbox{Tr}\; \mathsf{\Phi} \mathsf{\sigma}_{1} \mathsf{\Psi}
\right)^{2}
\end{equation}
with 
\begin{equation}
  \dirac = 
  \left( \begin{array}{cc} 
  \partial_{\xi} & 0 \\
  0 & \partial_{\eta}  
  \end{array} \right).
\end{equation}

It is easy to see that \eref{def:lagrA} resembles the Gross-Neveu Lagrangian 
\eref{def:LagrGNa}. The main difference is that the Lagrangian \eref{def:lagrA} 
is built of \emph{two} matrices, $\mathsf{\Phi}$ and $\mathsf{\Psi}$. 
In the following section we discuss questions related to 
complex/Hermitian conjugation and establish that there is a natural reduction 
which links $\mathsf{\Phi}$ and $\mathsf{\Psi}$.

\section{Involution. \label{sec:involution}}

It turns out that, if one works in the framework of the soliton ansatz 
used in this paper, the behavior of solutions under the complex/Hermitian 
conjugation is determined by whether the matrices $\mathsf{L}$ are real or 
imaginary. Indeed, it is not difficult to show that the condition
\begin{equation}
\label{inv:L}
  \mathsf{L}\hconj = - \epsilon \mathsf{L},
  \quad
  \epsilon = \pm 1
\end{equation}
implies 
\begin{equation}
  \mathsf{A}\hconj = \mathsf{A}, \qquad
  \mathsf{G}\hconj = \mathsf{G}.
\end{equation}
The requirement 
\begin{equation}
  \mathsf{E}\hconj
  = 
    \mathsf{E}^{-1}
\end{equation}
leads to the restrictions 
\begin{equation}
  z_{n}^{*} = \epsilon z_{n} 
  \qquad
  (n = 1, \, ... \, ,\Nz)
\end{equation}
and 
\begin{equation}
\label{inv:xi}
  \xi^{*} = - \epsilon \xi, 
  \qquad
  \eta^{*} = - \epsilon \eta 
\end{equation}
where $*$ stands for the complex conjugation. 

This results in 
\begin{equation}
  h^{*} = \epsilon h, 
\end{equation}
\begin{equation}
  \mathsf{\Phi} = \mathsf{\Psi}\hconj \mathsf{S} 
\end{equation}
where 
\begin{equation}
  \mathsf{S}
  = 
  \left(\begin{array}{cc}
    1 & 0 \\ 0 & \epsilon 
  \end{array}\right) 
\end{equation}
and hence
\begin{equation}
\label{def:lagrB}
  \Lagrangian
  = 
  4 h \, \mbox{Tr}\, 
    \mathsf{\Psi}\hconj \mathsf{S} \!\dirac \mathsf{\Psi}
  - \left( \mbox{Tr}\; 
      \mathsf{\Psi}\hconj 
      \mathsf{S} 
      \mathsf{\sigma}_{1} 
      \mathsf{\Psi} 
    \right)^{2}. 
\end{equation}

\bigskip

To conclude our analysis, we consider separately the cases  
$\epsilon = \pm 1$, and rewrite the Lagrangian \eref{def:lagrB} in 
terms of the Dirac matrices, 
\begin{equation}
  \dm{0} = \mathsf{\sigma}_{1}
\qquad
  \dm{1} = i \mathsf{\sigma}_{2}
\qquad
  \dm{5} = - \mathsf{\sigma}_{3}
\end{equation}
and the adjoint matrix $\bar{\mathsf{\Psi}}$ defined by 
\begin{equation}
  \bar{\mathsf{\Psi}} = \mathsf{\Psi\hconj} \dm{0}
\end{equation}

\subsubsection*{Gross-Neveu case ($\epsilon = 1$).}

To take into account the fact that in this case, as follows from \eref{inv:xi}, 
both $\xi$ and $\eta$ are pure imaginary we introduce two real variables, 
\begin{equation}
  \xi = i (t - x), \qquad \eta = i(t + x).
\end{equation}
Noting that $\mathsf{S}$ is the unit matrix and that 
$h$ is real we can introduce the real coupling constant
\begin{equation}
  g = \frac{ 1 }{ 2h }
\end{equation}
and rewrite the Lagrangian \eref{def:lagrB} 
(omitting an insignificant constant) as 
\begin{equation}
  \Lagrangian
  = 
  i \mbox{Tr}\; \mathsf{\bar{\Psi}} \!\dirac\mathsf{\Psi} 
  + g \left(\mbox{Tr}\; \mathsf{\bar{\Psi}} \mathsf{\Psi} \right)^{2}
\end{equation}
with 
\begin{equation}
\label{def:dirac-tx}
  \dirac
  = 
  \dm{0}\partial_{t} 
  + \dm{1}\partial_{x}, 
\end{equation}
which coincides with \eref{def:LagrGNa}.

\subsubsection*{ Case $\epsilon = -1$.}

In this case, both $\xi$ and $\eta$ are real, 
$\mathsf{S}=\mathsf{\sigma}_{3}$, 
the parameters $z_{n}$ and hence the constant $h$ are pure imaginary. 

Introducing $t$ and $x$ by 
\begin{equation}
  \xi = -t + x, \qquad \eta = t + x.
\end{equation}
and $g$ by 
\begin{equation}
  g = \frac{ i }{ 2h }
\end{equation}
(all of which are real), 
one can rewrite the Lagrangian \eref{def:lagrB} 
(again, omitting an insignificant constant) as 

\begin{equation}
\label{def:CGN}
  \Lagrangian
  = 
  i \mbox{Tr}\; \mathsf{\bar{\Psi}} \!\dirac\mathsf{\Psi} 
  + g \left(\mbox{Tr}\; 
        \mathsf{\bar{\Psi}} 
        \dm{5} 
        \mathsf{\Psi} 
      \right)^{2}, 
\end{equation}
with $\dirac$ defined in \eref{def:dirac-tx}.

\section{Solitons of the Gross-Neveu model. \label{sec:solitons} }

Here, we would like to collect the results related to the Gross-Neveu model.

As follows from \eref{inv:L} with $\epsilon = 1$, the matrix 
$\mathsf{L}$ is pure imaginary. Thus, we write it as 
\begin{equation} 
\label{sls:L}
  \mathsf{L} = 
  i \mathsf{M} = 
  i \mathop{\mbox{diag}}\left( 
      \mu_{1}, \, ... \, ,\mu_{\scriptscriptstyle\Ns} 
    \right) 
\end{equation}
with real $\mu_{m}$.
Equation \eref{eq:sy} leads to 
\begin{equation} 
  \mathsf{A}(t,x) = \biggl(
    C_{lm} \exp\left[ \omega_{l}(t,x) + \omega_{m}(t,x) \right]
  \biggr)_{\scriptscriptstyle l,m = 1, ..., \Ns}
\end{equation} 
where 
\begin{equation} 
\label{sls:omega}
  \omega_{m}(t,x) 
  = 
  \left(\mu_{m} - \mu_{m}^{-1}\right) t + 
  \left(\mu_{m} + \mu_{m}^{-1}\right) x
\end{equation} 
or 
\begin{equation} 
  \omega_{m}(t,x) = \frac{2}{1 - v_{m}^{2}} \, (x - v_{m} t ), 
  \quad
  v_{m} = \frac{ 1 - \mu_{m}^{2} }{ 1 + \mu_{m}^{2} }   
\end{equation} 
and $C_{lm}$ are constants given by 
\begin{equation} 
C_{lm} = -i \frac{ \k_{l}^{(0)} \b_{m}^{(0)} }{ (\mu_{l}+\mu_{m}) } 
\end{equation} 
with $\b_{m}^{(0)} = \b_{m}(0,0)$ and $\k_{m}^{(0)} = \k_{l}(0,0)$ 
playing the role of parameters of the presented solutions.

The columns of the matrix $\mathsf{\Psi}$ can be presented as 
\begin{equation} 
  \vec{\psi}_{n} = 
  e^{ i \Theta_{n} }
  \left(
  \begin{array}{l} 
    1 -
    \langle \b | \mathsf{G} (\mathsf{L} - z_{n})^{-1} | \k \rangle 
  \\[2mm]
    z_{n}^{-1} +
    \langle \b | 
      \mathsf{L}^{-1} \mathsf{G} (\mathsf{L} - z_{n})^{-1} 
    | \k \rangle 
  \end{array} 
  \right)
\end{equation}
where the phases $\Theta_{n} = \Theta_{n}(t,x)$ are given by 
\begin{equation} 
\label{sls:Omega}
  \Theta_{n}(t,x) 
  = 
  \left(z_{n} + z_{n}^{-1}\right) t + \left(z_{n} - z_{n}^{-1}\right) x
\end{equation} 
or 
\begin{equation} 
\label{sls:psi}
  \vec{\psi}_{n}(t,x) = 
  \frac{ e^{ i \Theta_{n}(t,x) } }
       { \det\left| z_{n} - i \mathsf{M} \right| } 
  \left(
  \begin{array}{l} 
  \det\left| z_{n} - i \mathsf{M}\mathsf{Y}(t,x) \right|
  \\[2mm]
  z_{n}^{-1} 
  \det\left| z_{n}\mathsf{Y}(t,x) - i \mathsf{M} \right|
  \end{array} 
  \right)
\end{equation}
where 
\begin{equation} 
\label{sls:Y}
  \mathsf{Y} = 
  \left( \mathsf{1} + \mathsf{A} \right)^{-1} 
  \left( \mathsf{1} - \mathsf{A} \right). 
\end{equation}

To summarize, formulae \eref{sls:psi} together with 
\eref{sls:L}--\eref{sls:omega}, \eref{sls:Omega} and \eref{sls:Y}
provide the $\Ns$-soliton solutions for the Gross-Neveu model.

The function $u_{o}$, which in this case can be presented as 
\begin{equation} 
\label{eq:uodet}
  u_{o} = \frac{\det\left| \mathsf{1} - \mathsf{A} \right|}
               {\det\left| \mathsf{1} + \mathsf{A} \right|},
\end{equation} 
satisfies the $\sinh$-Gordon equation
\begin{equation} 
\label{eq:SG}
  \frac{1}{4} \, \Box\, u_{o} = u_{o}^{-2} - u_{o}^{2} 
\end{equation} 
where $\Box = \partial_{tt} - \partial_{xx}$ 
(we prove these facts in Appendix \ref{app:tau}).

It is not difficult to obtain from \eref{sls:psi} the behavior of 
$\vec{\psi}_{n}$ in the asymptotic regions. For simplicity, we carry out this 
analysis under the following assumption:
\begin{equation} 
  \mu_{s} > 0, \quad s = 1, \, ... \, , \Ns.
\end{equation}

When $x \to -\infty$ with $t = \mbox{constant}$, 
all $\omega_{m}(t,x) \to -\infty$ which yields 
$\mathsf{Y}(t,x) \to \mathsf{1}$ and
\begin{equation} 
  \lim_{x \to -\infty} 
  e^{ -i \Theta_{n}(t,x) } \vec{\psi}_{n}(t,x) 
  = 
  \vec{\phi}_{n}^{\scriptscriptstyle-}
\end{equation} 
with 
\begin{equation} 
  \vec{\phi}_{n}^{\scriptscriptstyle-}
  =
  \left(\begin{array}{c} 1 \\ z_{n}^{-1} \end{array}\right). 
\end{equation}
In a similar way one arrives at  
\begin{equation} 
  \lim_{x \to +\infty} 
  e^{ -i \Theta_{n}(t,x) } \vec{\psi}_{n}(t,x) 
  =
  \vec{\phi}_{n}^{\scriptscriptstyle+}
\end{equation} 
with 
\begin{equation} 
  \vec{\phi}_{n}^{\scriptscriptstyle+} =
  e^{i\delta_{n}}
  \left(\begin{array}{c} 1 \\ (-)^{\Ns} z_{n}^{-1} \end{array}\right) 
\end{equation}
and 
\begin{equation} 
 \delta_{n} 
 = 
 2 \sum_{s=1}^{\Ns} \arg\left( z_{n} + i\mu_{s} \right). 
\end{equation} 

The limiting values of the condensate function $S$, 
\begin{equation} 
  S = 
  \mbox{Tr}\,\bar{\mathsf{\Psi}}\mathsf{\Psi} =
  \sum_{n=1}^{\Nz} \bar{\vec{\psi}}_{n}\vec{\psi}_{n},   
\label{def:S}
\end{equation} 
are given by 
\begin{equation} 
  S(t,x) \to
  \left\{
  \begin{array}{rl}
  1/g         & \mbox{as}\; x \to -\infty \\ 
  (-)^{\Ns}/g & \mbox{as}\; x \to +\infty 
  \end{array}
  \right.
\end{equation} 
while the fermion density $Q$, 
\begin{equation} 
\label{def:Q}
  Q 
  =
  \mbox{Tr}\,\mathsf{\Psi}\hconj\mathsf{\Psi} 
  = 
  \sum_{n=1}^{\Nz} \vec{\psi}_{n}\hconj\vec{\psi}_{n}, 
\end{equation} 
satisfies 
\begin{equation} 
  \lim_{x \to \pm\infty} Q(t,x) = 
  Q^{\scriptscriptstyle\infty} = \mbox{constant} 
\end{equation} 
where 
\begin{equation} 
\label{def:Qinf}
  Q^{\scriptscriptstyle\infty} =
  \Nz + \sum_{n=1}^{\Nz} z_{n}^{-2} 
\end{equation}

For the one-soliton solution 
$\Ns=1$, the matrix $\mathsf{L}$ is scalar, 
$\mathsf{L} = i\mu$ (we drop the subscript $1$), 
and the one-soliton solution is characterized, 
except for the real set $\{z_{1}, \, ... \, ,z_{\Nz} \}$, by one velocity $v$, 
$v = ( 1 - \mu^{2} )/( 1 + \mu^{2} )$,
and one constant $C_{11}$, which without loss of generality can be set equal to 
unity, $C_{11}=1$. The matrix $\mathsf{Y}$ becomes 
\begin{equation} 
  \mathsf{Y} = - \tanh\omega, 
\end{equation} 
where 
\begin{equation} 
\omega(t,x) = \frac{2}{1 - v^{2}} \left( x - v t \right).
\end{equation} 
Equation \eref{sls:psi} can be rewritten as 
\begin{equation} 
\label{sls:one}
  \vec{\psi}_{n}(t,x) = 
  \frac{1}{2} e^{ i \Theta_{n}(t,x) }  
  \bigl[  
    \vec{\phi}_{n}^{\scriptscriptstyle+} + 
    \vec{\phi}_{n}^{\scriptscriptstyle-} 
    + 
    \left( \vec{\phi}_{n}^{\scriptscriptstyle+} 
         - \vec{\phi}_{n}^{\scriptscriptstyle-} \right)
    \tanh\omega 
  \bigr] 
\end{equation}
or
\begin{equation} 
  \vec{\psi}_{n}(t,x) = 
  \frac{ e^{ i \Theta_{n}(t,x) } }{ 2\cosh\omega } 
  \left[ 
    e^{ - \omega } \vec{\phi}_{n}^{\scriptscriptstyle-} + 
    e^{   \omega } \vec{\phi}_{n}^{\scriptscriptstyle+} 
  \right], 
\end{equation}
where $\vec{\phi}_{n}^{\scriptscriptstyle\pm}$ are the limits of 
$e^{-i\Theta_{n}}\vec{\psi}_{n}$ defined earlier and given by  
\begin{equation} 
  \vec{\phi}_{n}^{\scriptscriptstyle-} = 
  \left(\begin{array}{c} 1 \\ z_{n}^{-1} \end{array}\right),
  \qquad
  \vec{\phi}_{n}^{\scriptscriptstyle+} = 
  e^{ i \delta_{n} }
  \left(\begin{array}{c} 1 \\ -z_{n}^{-1} \end{array}\right),
\end{equation}
with 
\begin{equation} 
 \delta_{n} 
 = 
 2 \arg\left( z_{n} + i\sqrt{\frac{1-v}{1+v}} \right). 
\end{equation} 

The distribution of the condensate $S$, 
which is defined in \eref{def:S}, is given by 
\begin{equation} 
\label{sls:S-one}
  S = - \frac{1}{g} \,\tanh\omega, 
\end{equation} 
while the fermion density $Q$, defined in \eref{def:Q}, can be presented as 
\begin{equation} 
  Q
  = 
  Q^{\scriptscriptstyle\infty} - \frac{A}{\cosh^{2}\omega} 
\end{equation} 
with $Q^{\scriptscriptstyle\infty}$ being defined in \eref{def:Qinf} and 
\begin{equation} 
  A 
  = 
  (1 + \mu^{2}) 
  \sum_{n=1}^{\Nz} \frac{ 1 }{ z_{n}^{2} + \mu^{2} }. 
\end{equation} 

\begin{figure}
\includegraphics{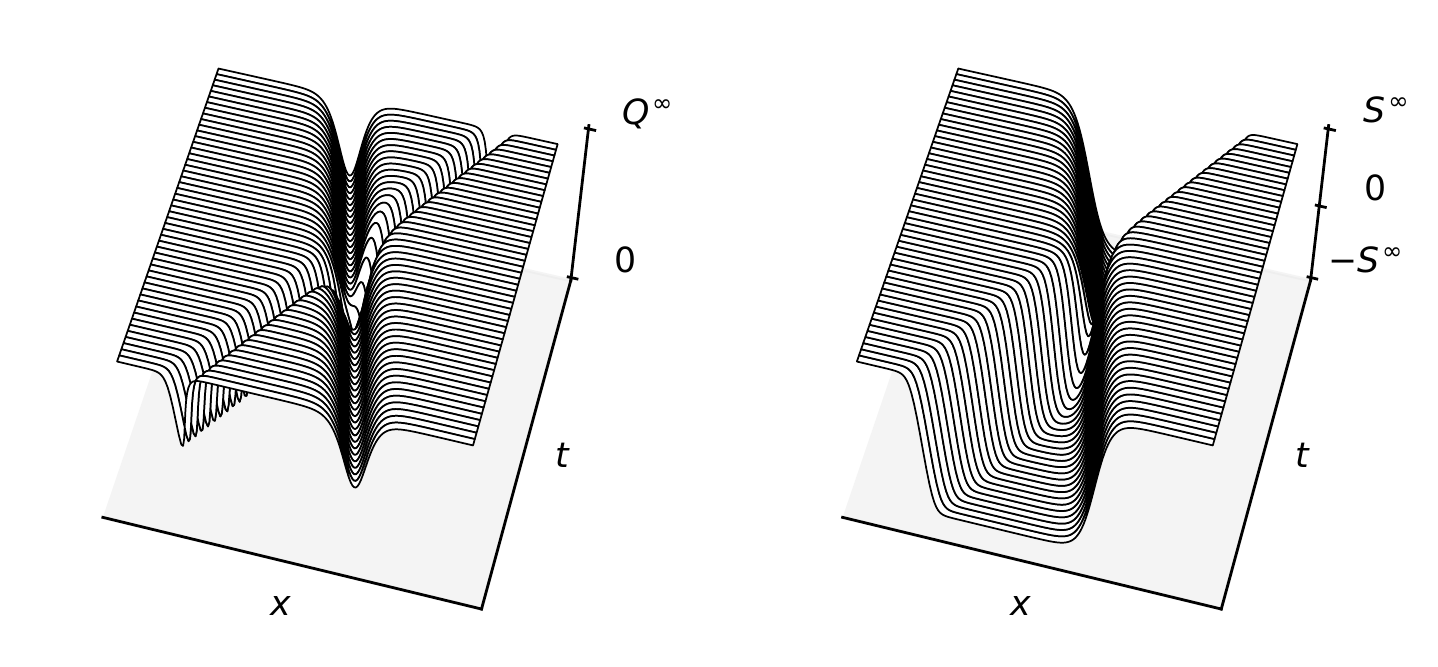}
\caption{\label{fig1}
2-soliton solutions: 
fermion density $Q(t,x)$ (left pane) and condensate $S(t,x)$ (right pane). 
The parameters of the solution are given by 
$\mathsf{L} = i \, \mbox{diag}\,(0.2,1.5)$, 
$\langle \b(0,0) | = e^{i\pi/6} (1,1)$, 
$| \k(0,0) \rangle = e^{i\pi/3} (1,1)\transposed$ 
and $\myset{Z} = \{1\}$.
The coordinate ranges are given by $t \in (- 3.5,3.5)$ and $x \in (- 5.0,5.0)$. 
}
\end{figure}

\begin{figure}
\includegraphics{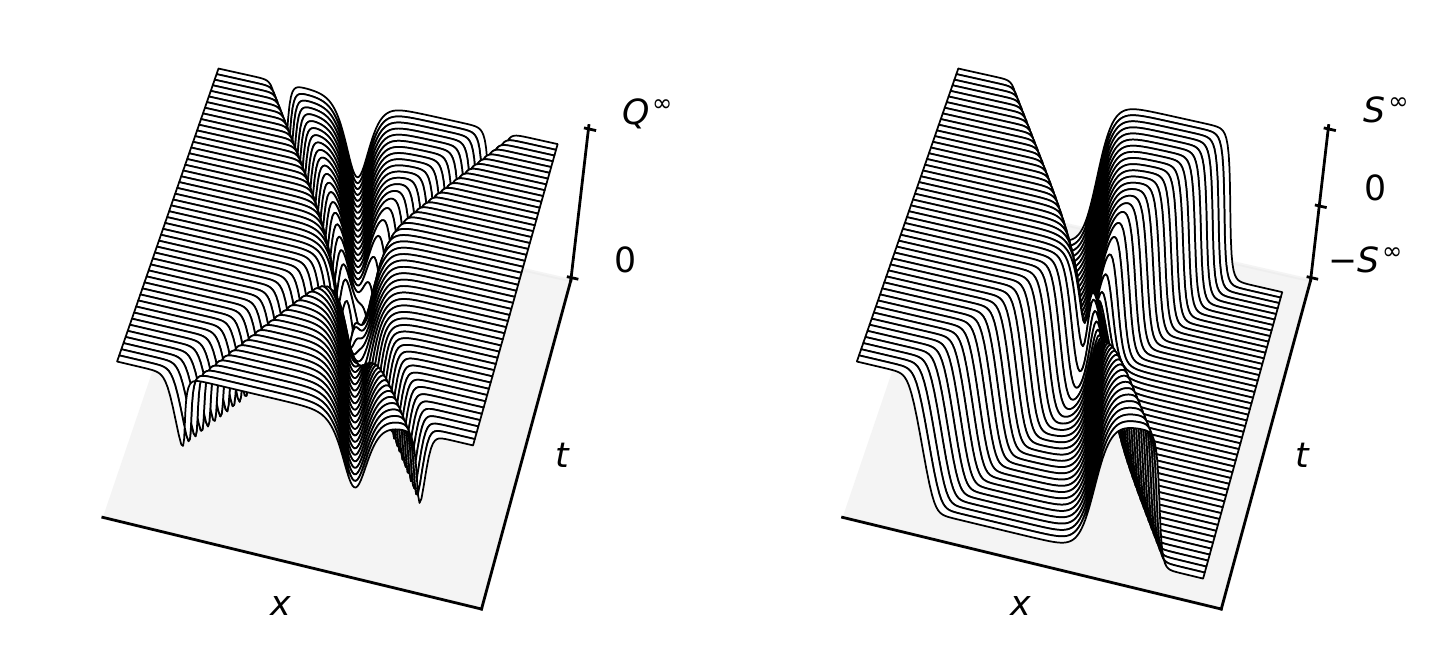}
\caption{\label{fig2}
3-soliton solutions: 
fermion density $Q(t,x)$ (left pane) and condensate $S(t,x)$ (right pane).
The parameters of the solution are given by 
$\mathsf{L} = i \, \mbox{diag}\,(0.2,1.5,6.0)$, 
$\langle \b(0,0) | = e^{i\pi/6} (1,1,1)$, 
$| \k(0,0) \rangle = e^{i\pi/3} (1,1,1)\transposed$ 
and $\myset{Z} = \{1\}$.
The coordinate ranges are given by $t \in (- 3.5,3.5)$ and $x \in (- 5.0,5.0)$. 
}
\end{figure}

\bigskip
Considering the more complex solutions,
we present examples of two- and three-soliton solutions in 
figures \ref{fig1} and \ref{fig2}.

\section{Discussion.}

To derive the solitons of the Gross-Neveu model we used rather standard 
technique from the theory of integrable systems. 
The Cauchy matrix approach, which appeared in 1980s as an alternative to the 
inverse scattering transform, was subsequently modified to become one of the 
easiest way to derive explicit solutions for integrable nonlinear equations. 
In this paper, and in many others, it is used 
just like an ansatz, which, if compared, for example, with the inverse 
scattering transform, is more straightforward, not restricted by imposing 
some boundary conditions beforehand, and rather flexible 
(see, e.g., \cite{V14}). Even in the framework of this paper one can note that 
our ansatz, with slight modifications, leads to solutions for both the 
Gross-Neveu model and its $\dm5$ variant \eref{def:CGN}.
Clearly, it has its limitations. 
The soliton ansatz of this paper, that in context of other integrable models 
leads to `general' N-soliton solutions, in the case of the Gross-Neveu 
equations provides less than one might anticipate. The solutions presented 
above belong to the so-called type I (the simplest) class, according to the 
classification of \cite{KT10}.
Indeed, if we take notice of the $z_{n}$-dependence, the 
condensate function $S$ can be presented as 
\begin{equation} 
  S(t,x,\myset{Z}) = 2 h(\myset{Z}) u_{o}(t,x)
\end{equation} 
which leads to 
\begin{equation} 
  \bar{\vec{\psi}}_{n}\vec{\psi}_{n} 
  = 
  \lambda_{n} S
\end{equation} 
where $\lambda_{n}$ is a \emph{constant}, given by 
$\lambda_{n} = z_{n}^{-1} \left/ \sum_{m=1}^{\Nz} z_{m}^{-1} \right.$.
Thus, in our attempt to derive the $\Nz$-flavor solitons we have actually 
obtained some kind of direct sum of $\Nz=1$ solitons 
(up to the linear global mixing $\mathsf{\Psi} \to \mathsf{\Psi}\mathsf{U}$ 
where $\mathsf{U}$ is a \emph{constant} unitary matrix). 
This means that to find less trivial $\Nz$-flavor solutions 
(even if we restrict ourselves to the classical Gross-Neveu model with 
finite $\Nz$, i.e. without continuous constituent) we have to go beyond 
the soliton ansatz used above. However, this very important question is 
outside of the scope of this paper. 

The last question we would like to mention is the question of terms. 
In the theory of integrable systems solutions like ones presented in this paper 
are usually called `solitons'  or, more precisely, `dark solitons' where the 
word `dark' indicates that they are solitons which satisfy constant non-zero 
boundary conditions. In the field theory, the more widely used term is `kink'. 
If one looks at the one-soliton solution \eref{sls:one} (or \eref{sls:S-one}), 
then there is no discordance: the $\tanh$-function is what is usually 
associated with a kink. 
However, in the situation with two (or any even $\Ns$) solitons 
the asymptotic behavior of both the condensate $S$ and the fermion density $Q$ 
differs from the kink-, or $\Ns$-kink-like, one 
(see, for example, figure~\ref{fig1}).

\appendix

\section{Derivation of \eref{eq:sp}. \label{app:products} }

From the definitions of 
$\mathsf{B}_{\myset{Z}}$ and 
$\mathsf{K}_{\myset{Z}}$,  
one can easily derive
\begin{equation}
  \langle 1_{\myset{Z}} | \mathsf{B}_{\myset{Z}}
  = 
  \langle \b | \mathsf{\hat{B}}_{\myset{Z}}, 
  \quad
  \mathsf{K}_{\myset{Z}} | 1_{\myset{Z}} \rangle
  = 
  \mathsf{\hat{K}}_{\myset{Z}} | \k \rangle
\end{equation}
and 
\begin{eqnarray}
  \langle 1_{\myset{Z}} | \mathsf{D}_{\myset{Z}}^{-1} 
  \mathsf{B}_{\myset{Z}}
  & = & 
  \langle \b_{o} | \left(h +\mathsf{\hat{B}}_{\myset{Z}} \right), 
\\
  \mathsf{K}_{\myset{Z}} 
  \mathsf{D}_{\myset{Z}}^{-1} | 1_{\myset{Z}} \rangle
  & = & 
  \left( h + \mathsf{\hat{K}}_{\myset{Z}} \right) | \k_{o} \rangle
\end{eqnarray} 
where $h$ is defined in \eref{def:h} and  
\begin{equation}
  \mathsf{\hat{B}}_{\myset{Z}}
  = 
  \sum_{z \in  \myset{Z}} \mathsf{R}_{z}^{-1}, 
  \qquad
  \mathsf{\hat{K}}_{\myset{Z}}
  = 
  \sum_{z \in  \myset{Z}} \mathsf{L}_{z}^{-1}. 
\end{equation}
These equations, together with the definitions \eref{def:bu} and \eref{def:ku}, 
lead to 
\begin{equation}
\label{app:sp}
  \begin{array}{l}
  \langle \bu1_{\myset{Z}} | \ku2_{\myset{Z}} \rangle
  = 
  h u_{o} - \langle \b | \mathsf{F}_{\myset{Z}} | \k_{o} \rangle
  \\
  \langle \bu2_{\myset{Z}} | \ku1_{\myset{Z}} \rangle
  = 
  h v_{o} - \langle \b_{o} | \mathsf{F}_{\myset{Z}} | \k \rangle
  \end{array}
\end{equation}
with 
\begin{equation}
  \mathsf{F}_{\myset{Z}}
  = 
    \mathsf{G} \mathsf{K}_{\myset{Z}} \mathsf{B}_{\myset{Z}} \mathsf{G}
  - \mathsf{\hat{B}}_{\myset{Z}} \mathsf{G}
  + \mathsf{G} \mathsf{\hat{K}}_{\myset{Z}}. 
\end{equation}
To simplify this expression, one should note that 

\begin{equation}
  \mathsf{K}_{\myset{Z}} \mathsf{B}_{\myset{Z}}
  = 
  \sum_{n=1}^{\Nz} | \k_{z_{n}} \rangle \langle \b_{z_{n}} |
\end{equation}
where 
\begin{equation}
  \langle \b_{z} | = \langle \b | \left( \mathsf{R} - z \right)^{-1}, 
  \quad
  | \k_{z} \rangle = \left( \mathsf{L} - z  \right)^{-1} | \k \rangle,
\end{equation}
which, together with \eref{eq:sy}, leads to 
\begin{equation}
    \mathsf{K}_{\myset{Z}} \mathsf{B}_{\myset{Z}}
  = 
    \mathsf{A} \mathsf{\hat{B}}_{\myset{Z}}
  - \mathsf{\hat{K}}_{\myset{Z}} \mathsf{A}
\end{equation}
and, finally, to 
\begin{equation}
\label{app:F}
  \mathsf{F}_{\myset{Z}}
  = 
  \mathsf{G} 
  \left( \mathsf{\hat{K}}_{\myset{Z}} - \mathsf{\hat{B}}_{\myset{Z}} \right) 
  \mathsf{G}.
\end{equation}
Clearly, equations \eref{app:sp} and \eref{app:F} coincide with 
\eref{eq:sp} and \eref{def:F} with 
$ \mathsf{\hat{F}}_{\myset{Z}} = 
\mathsf{\hat{K}}_{\myset{Z}} - \mathsf{\hat{B}}_{\myset{Z}}$.

\section{Proof of \eref{inv:uv} and \eref{inv:FZ}. \label{app:reduction} }

First, one has to note that the matrix $\mathsf{T}$ that links 
$| \k \rangle$ and $\langle \b |$, 
\begin{equation}
  | \k \rangle = \mathsf{T} | \b\transposed \rangle  
\end{equation}
where 
$|\b\transposed\rangle = ( \langle \b | )\transposed$, 
and which is given by 
\begin{equation}
  \mathsf{T} =  
  \mbox{diag} \left( ... \, , \k_{n}/\b_{n}, \, ... \right),
\end{equation}
as follows from equations \eref{def:doib} and \eref{def:doik} together with 
the restriction \eref{eq:LR}, 
does not depend on $\xi$ or $\eta$.

From \eref{eq:sy} one can easily obtain
$\mathsf{A}\transposed = \mathsf{T}^{-1}\mathsf{A}\mathsf{T}$, 
which holds for all $\xi$ and $\eta$ and which implies
$\mathsf{G}\transposed = \mathsf{T}^{-1}\mathsf{G}\mathsf{T}$.
Noting also that 
$| \k_{o} \rangle 
= - \mathsf{T} | \b_{o}\transposed \rangle$, 
where 
$| \b_{o}\transposed \rangle =(\langle \b_{o} |)\transposed$, 
one can consequently obtain from \eref{def:uv}
\begin{eqnarray}
  u_{o} 
  & = & 
  1 - \langle \b | \mathsf{G} | \k_{o} \rangle 
  \nonumber\\ & = & 
  1 + \langle \b | \mathsf{G}\mathsf{T} | \b_{o}\transposed \rangle 
  \nonumber\\ & = & 
  1 + \langle \b_{o} | \mathsf{T}\mathsf{G}\transposed | \b\transposed \rangle 
  \nonumber\\ & = & 
  1 + \langle \b_{o} | \mathsf{G} | \k \rangle 
  \\ & = & 
  v_{o} 
\end{eqnarray}
which proves \eref{inv:uv}.

In a similar way, noting that 
\begin{equation} 
  \left( | \k_{o} \rangle \langle \b | \right)\transposed 
  = 
  - | \b\transposed \rangle \langle \b_{o} | \mathsf{T} 
  = 
  - \mathsf{T}^{-1} | \k \rangle \langle \b_{o} | \mathsf{T}, 
\end{equation} 
and that the matrix $\mathsf{\hat{F}}_{\myset{Z}}$ in the 
definition \eref{def:F} is diagonal (and hence commutes with $\mathsf{T}$)
one can obtain 
\begin{equation} 
  \mathsf{F}_{\myset{Z}}\transposed 
  = 
  \mathsf{T}^{-1} \mathsf{F}_{\myset{Z}} \mathsf{T} 
\end{equation} 
and then 
\begin{eqnarray}
  \langle \b | \mathsf{F}_{\myset{Z}} | \k_{o} \rangle 
  & = &
  \mbox{Tr}\; 
    \mathsf{F}_{\myset{Z}} | \k_{o} \rangle \langle \b | 
  \nonumber\\ & = & 
  \mbox{Tr}\; 
    (|\k_{o}\rangle \langle\b|)\transposed \mathsf{F}_{\myset{Z}}\transposed 
  \nonumber\\ & = & 
  - \mbox{Tr}\; 
    \mathsf{T}^{-1} | \k \rangle \langle \b_{o} | \mathsf{T} \cdot 
    \mathsf{T}^{-1} \mathsf{F}_{\myset{Z}}\mathsf{T} 
  \nonumber\\ & = & 
  - \mbox{Tr}\; 
    | \k \rangle \langle \b_{o} | 
    \mathsf{F}_{\myset{Z}} 
  \nonumber\\ & = & 
  - \langle \b_{o} | \mathsf{F}_{\myset{Z}} | \k_ \rangle 
\end{eqnarray}
which proves \eref{inv:FZ}.

\section{Tau-functions. \label{app:tau} }

The $\tau$-functions of the Gross-Neveu model can be defined as 
\begin{equation} 
  \tau_{\pm} = \det| \mathsf{1} \pm \mathsf{A} |.
\end{equation} 
From equation \eref{eq:sy} with $\mathsf{R} = -\mathsf{L}$ and the definition 
\eref{def:boko} of $| \k_{o} \rangle$ one can easily obtain
\begin{equation} 
  \mathsf{1} + \mathsf{A} - | \k_{o} \rangle \langle \b | = 
  \mathsf{1} - \mathsf{L}^{-1} \mathsf{A} \mathsf{L} 
\end{equation} 
and then 
\begin{equation} 
  \mathsf{1} - \mathsf{G} | \k_{o} \rangle \langle \b | = 
  \mathsf{G} \mathsf{L}^{-1}
  \left( \mathsf{1} -  \mathsf{A} \right) \mathsf{L} 
\end{equation} 
Taking the determinant of the last equation, using the identity 
$\det( \mathsf{1} + | u \rangle \langle v |) = 1 + \langle v || u \rangle$ 
and noting that 
$\det\mathsf{G} = 1 / \det| \mathsf{1} + \mathsf{A} |$ 
one arrives at 
\begin{equation} 
\label{tau:uo}
  u_{o} = \tau_{-} / \tau_{+}
\end{equation} 
which is \eref{eq:uodet}. 

The derivatives of $\tau_{\pm}$ are given by 
\begin{equation} 
  \partial_{\xi} \ln\tau_{\pm} = 
  \pm \langle \b_{o} | \mathsf{G}_{\pm} | \k_{o} \rangle, 
\qquad 
  \partial_{\eta} \ln\tau_{\pm} = 
  \mp \langle \b | \mathsf{G}_{\pm} | \k\rangle, 
\end{equation} 
where
\begin{equation} 
  \mathsf{G}_{\pm} = (\mathsf{1} \pm \mathsf{A})^{-1} 
\end{equation} 
from which one can derive 
\begin{equation} 
  \partial_{\xi\eta} \ln\tau_{\pm} = 1 - (\tau_{\mp}/\tau_{\pm})^{2} 
\end{equation} 
and, taking into account \eref{tau:uo},  
\begin{equation} 
  \partial_{\xi\eta} \ln u_{0} = u_{o}^{2} - u_{o}^{-2} 
\end{equation} 
which is nothing but \eref{eq:SG}.


\end{document}